\documentclass[aps,pra,
amsmath,
amssymb,showpacs,showkeys,reprint]{revtex4-1}

\usepackage{graphicx}
\usepackage{epstopdf}
\usepackage{upgreek}
\usepackage{float}

\begin{document}
\preprint{APS/123-QED}


\title{High Resolution Imaging and Optical Control of Bose-Einstein Condensates in an Atom Chip Magnetic Trap}


\author{Evan A. Salim}
 \altaffiliation[Current address: ]{ColdQuanta Inc., 1600 Range Street, Suite 103, Boulder, CO 80301}
\author{Seth C. Caliga}
\author{Jonathan B. Pfeiffer}
\author{Dana Z. Anderson}
\affiliation{Department of Physics, University of Colorado, and The JILA Institute, University of Colorado and National Institute for Standards and Technology, Boulder CO 80309-0440}


\date{\today}

\begin{abstract}
A high-resolution projection and imaging system for ultracold atoms is implemented using a compound silicon and glass atom chip.  The atom chip is metalized to enable magnetic trapping while glass regions enable high numerical aperture optical access to atoms residing in the magnetic trap about $100~\upmu$m below the chip surface.  The atom chip serves as a wall of the vacuum system, which enables the use of commercial microscope components for projection and imaging.  Holographically generated light patterns are used to optically slice a cigar-shaped magnetic trap into separate regions; this has been used to simultaneously generate up to four Bose-condensates.  Using fluorescence techniques we have demonstrated in-trap imaging resolution down to $2.5~\upmu\text{m}$.
\end{abstract}


\maketitle
High-resolution optical control of ultracold atoms has led to stunning experimental progress in the study and understanding of strongly correlated quantum systems such as the superfluid to Mott insulator transition~\cite{Bakr2009,Sherson2010} and antiferromagnetism~\cite{Simon2011}. Much of the experimental progress has been through the use of optical lattices~\cite{Bakr2009,Sherson2010,Simon2011,Nelson2007} formed by interfering far detuned laser beams. Imaging in such systems~\cite{Bakr2009,Sherson2010,Simon2011,Nelson2007,Bucker2009} has demonstrated sensitivity down to the single atom level.  More recently, high optical numerical aperture (NA) imaging and projection systems have also demonstrated novel trapping potentials such as toroids~\cite{Henderson2009} and ring lattices~\cite{Zimmermann2011}.  These projection systems achieve nearly arbitrary potential landscapes by generating optical patterns using either spatial light modulators~\cite{Brandt2011} or acousto-optic devices~\cite{Henderson2009,Zimmermann2011}. A compelling aspect of projected optical potentials as compared to interferometric approaches is the control it affords over atomic ensembles with high spatial \textit{bandwidth}, not simply high spatial frequency. This means that atomic structures of substantially different scales can coexist and be made to interact. For example, atomtronic analogues~\cite{Stickney2007} to semiconductor materials and devices may be realized by simultaneously manipulating both fine and coarse features in the potential landscape, with correspondingly small and large numbers of atoms.  In this work we introduce an atom chip system that enables simultaneous magnetic and optical control over an ultracold atomic ensemble with high spatial bandwidth and high-resolution imaging. 
The atom chip utilizes a compound substrate that is comprised of co-planar regions of glass and silicon. Through standard lithographic and metalization techniques, both glass and silicon can be simultaneously metalized to create the desired conductor pattern. This allows us to impose current-generated magnetic fields, in conjunction with external magnetic bias fields, to trap atoms below either silicon or glass regions of the chip

\begin{figure}[H]
\includegraphics{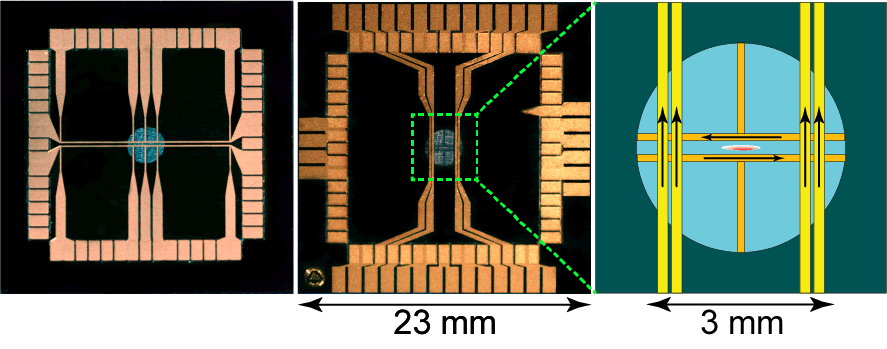}
\caption{\label{fig:Chip} 
Atom window chip -- (Left) Vacuum side metalization, with the main guide wires and intersecting wires for the ``T"-Trap.  UHV compatible electrical feedthroughs along the edge allow connection to in vacuo conductors. (Center) ``H"-wires on the ambient side and the pads along the perimeter, where connection is made between the chip and control electronics. (Right) Close-up of the window region and the current directions for the imaging waveguide.}  
\end{figure}

\noindent surface. The atom chip utilized in this work is shown in Fig.~\ref{fig:Chip}, with the vacuum and ambient sides shown in Fig.~\ref{fig:Chip} (left) and (center), respectively. 
The chip is $23~\text{mm}\times23~\text{mm}$,  $420~\upmu\text{m}$ thick, incorporates a 3 mm polished glass region in the center, and supports 60 ultra-high vacuum compatible electrical feedthroughs. Additional mechanical support is added to the ambient side of the chip on the area surrounding the window by bonding a 1 mm thick silicon backing ring to the atom chip in order to reduce bowing of the chip due to ambient pressure. The metalization pattern was designed utilizing LiveAtom~\cite{*[{Boulder Labs Inc, 7105 LaVista Pl. Suite 200, Niwot, CO 80503.  www.boulderlabs.com}]  [{}] Liveatom}. The atom chip forms the wall of a self-contained miniature double-MOT vacuum cell, the details and operation of which are described in greater detail elsewhere~\cite{Farkas2010,Salim2011}. The location of the trap in which we image the atoms is typically tens to hundreds of microns below the glass portion of the chip surface. In this configuration the NA of the imaging system is in principle limited by the edges of the chip window. This gives the system a potential NA up to 0.9 provided that the atoms are sufficiently close to the atom chip surface, while keeping all of the critical optical elements outside of the vacuum chamber. Typically, such high NAs would require custom in vacuo optics~\cite{Bakr2009,Sherson2010,Simon2011}. 

\begin{figure}[floatfix]
\includegraphics{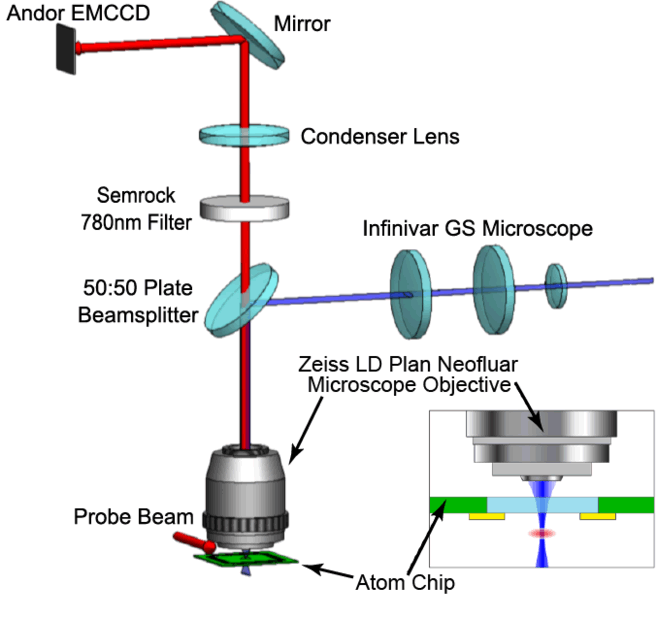}
\caption{\label{fig:Microscope} Schematic of the microscope system above the atom chip.  }  
\end{figure}

\begin{figure}[H]
\includegraphics{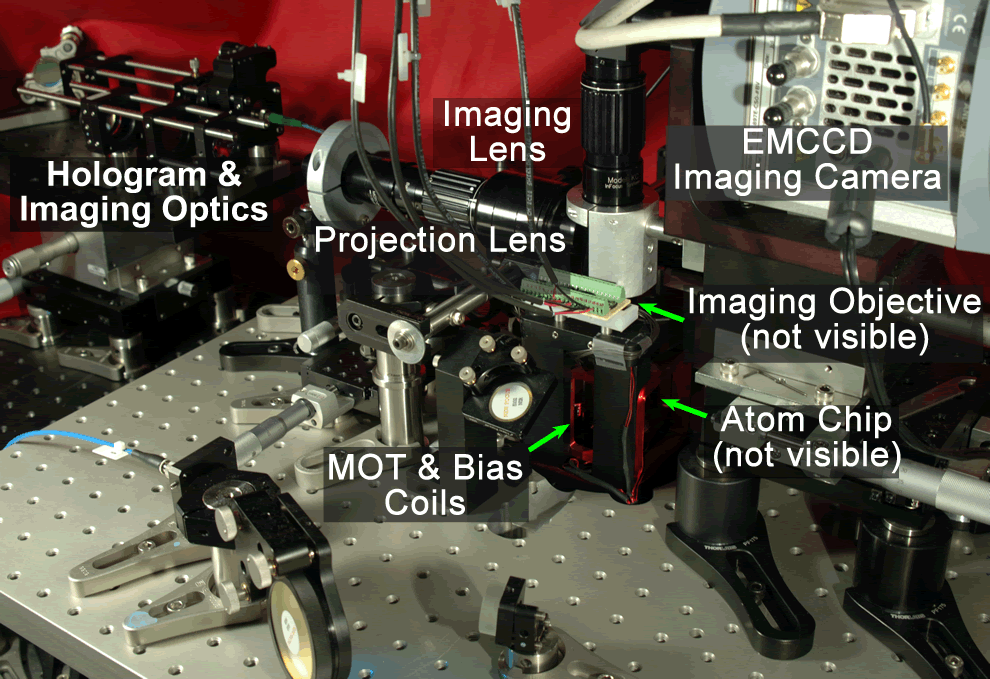}
\caption{\label{fig:Apparatus} High-resolution projection and imaging system.  The majority of 2D and 3D MOT optics lie beneath the 2' $\times$ 2' optical breadboard.  The atom chip vacuum cell protrudes through a hole in the breadboard. The red 3D MOT and transport coils that surround the upper cell chamber are just visible right of center in the photograph.  }  
\end{figure}

A microscope, (see Fig.~\ref{fig:Microscope}), is mounted directly over the atom chip with an integrated 50:50 beamsplitter to allow simultaneous high spatial bandwidth projection of optical potentials and high-resolution imaging of the atoms. In contrast to other experiments~\cite{Bakr2009,Sherson2010,Simon2011,Nelson2007,Bucker2009,Henderson2009,Zimmermann2011,Brandt2011}, our microscope system is made entirely from commercial components. Since the window chip is only $420~\upmu$m thick, commercially available microscope objectives with cover glass correction can be used without modification. We use a Zeiss LD Plan Neofluar objective, which has a magnification of 40X and a NA of 0.6. During the imaging process the light collected by the microscope objective is imaged through one port of the beamsplitter onto an Andor Ixon DU897, which is an electron multiplying CCD (EMCCD) camera, using an Infinity Photo-Optical KC series lens tube. The input to the other port of the beamsplitter is an Infinity Photo-Optical Infinivar GS infinity-corrected continuously-focusable microscope, which serves as the projection system for our optical potentials. The combination of the Infinivar GS and the Zeiss 40X objective images an object to the focal plane of the objective lens. To prevent the optical potential light from saturating the EMCCD a 780 nm notch filter that has a neutral density (ND) of 6 at the projection wavelength is placed on the camera side of the beamsplitter. The imaging and projection assembly, as well as the optics used for BEC production, are shown in Fig.~\ref{fig:Apparatus}. The majority of the 3D MOT optics are mounted on the underside of a custom 2' $\times$ 2' aluminum breadboard, leaving much of the top surface available for mounting components and instruments for future experiments. The result is a compact, robust platform for carrying out hybrid magnetic and optical

\noindent projection experiments on ultracold atomic ensembles.

\begin{figure}[floatfix]
\includegraphics{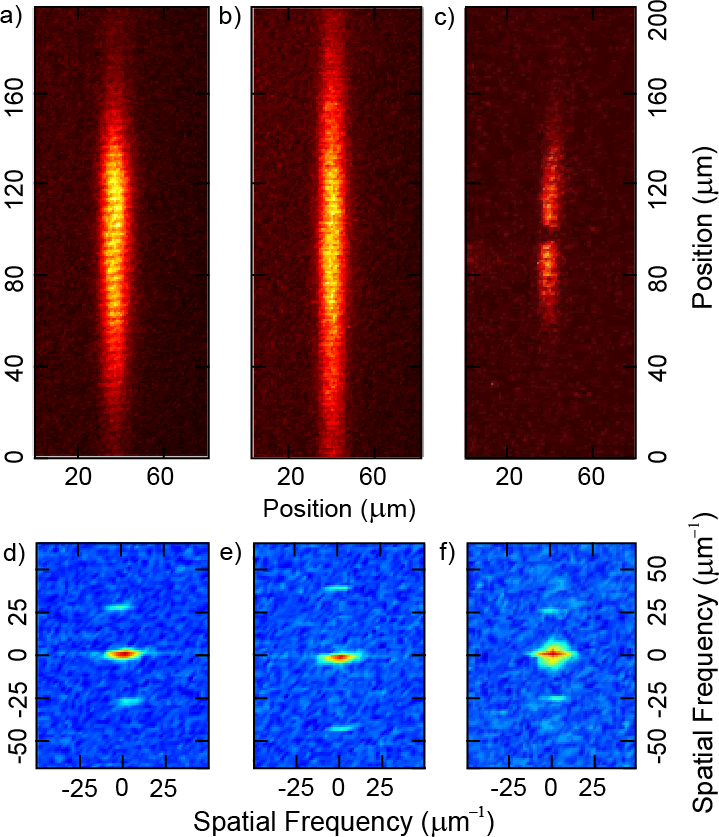}
\caption{\label{fig:SingleBarrier} Pictures of in-trap Images: (a) through (c) are five shot accumulations from the EMCCD of atoms in the magnetic imaging waveguide. Here (a) and (b) show the fluorescence of atoms in an interfering optical field with fringe spacing of 5 and $2.5~\upmu$m spacing, respectively. Splitting of a BEC with a $3~\upmu$m thick sheet of blue detuned light is shown in (c).  Fourier transforms of each image are shown directly below in (d) through (f).}  
\end{figure}

\begin{figure}[H]
\includegraphics{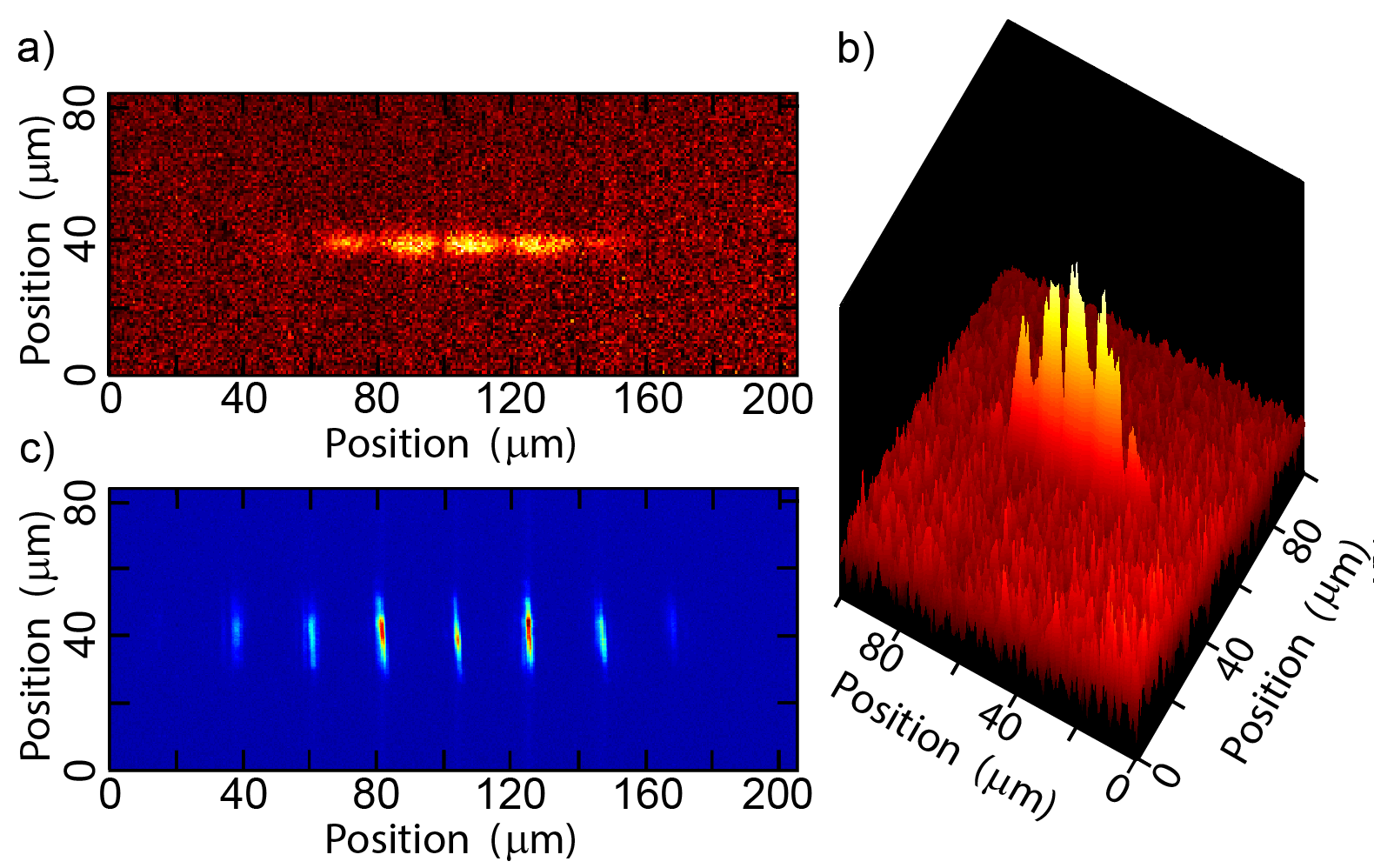}
\caption{\label{fig:MultiBarrier} Pictures of high contrast optical potentials and in-trap imaging: A 1D-Array of BECs is created, (a) and (b), by projecting a holographically generated array of blue detuned optical sheet barriers, (c), during RF evaporative cooling in imaging trap.}  
\end{figure}

To produce high-resolution images of ultracold atoms, two chip trap configurations are utilized in sequence.  First, atoms are adiabatically loaded into a tight dimple trap located $180~\upmu$m below the junction of the ``T"- shaped wires on the vacuum side of the chip using a pair of quadrupole coils. The dimple trap is generated by running 3 A in the main guide-wire and 0.35 A along the perpendicular T-wire, with bias fields in the $\vec{x}$ and $\vec{y}$ directions.  The trap frequencies for this trap are $140~\text{Hz}\times1.6~\text{kHz}\times1.8~\text{kHz}$.  Additional depth is provided in the loose trap direction by running 3.25 A in two pairs of parallel ``H"- shaped wires on the ambient side of the chip.  In this dimple trap, the atoms are cooled via forced radio-frequency (RF) evaporation over 1.6 seconds, in a similar manner as described in Ref.~\cite{Farkas2010}.  Once cooled to just above the transition temperature, the atoms are transferred adiabatically to an imaging trap located directly below the center of the window surface by ramping the chip wire currents and external bias fields. This trap is located $100~\upmu$m below the surface of the window with calculated trap frequencies of $67~\text{Hz}\times1.7~\text{kHz}\times1.7~\text{kHz}$. The imaging trap is generated using the current configuration shown in Fig.~\ref{fig:Chip} (right) where 2 A is run in opposite directions through the parallel guide-wires on the vacuum side and 1.75 A in the H-wire pairs, with bias fields in the $\vec{x}$ and $-\vec{z}$ directions.  The optical potential is then turned on adiabatically by ramping up the intensity in the desired trapping landscape.  The blue detuned light used for our optical potential is generated by a free running Coherent 899 Ti:sapphire ring laser operating at 760 nm.  Two potential landscapes are used, a single sheet barrier with FWHM dimensions of $3~\upmu$m by $20~\upmu$m and a 1D array of barriers with the same dimensions. These are shown in Fig.~\ref{fig:SingleBarrier} and Fig.~\ref{fig:MultiBarrier}. Additional RF evaporative cooling can be performed in the imaging trap either before or after raising the optical potential to cool the atoms to degeneracy.  

With the magnetic trap still on, a $400~\upmu$W, 7.5 mm 1/e diameter Gaussian beam of near-resonant probe light is pulsed for $20~\upmu$s as shown in Fig.~\ref{fig:Microscope}. Fluorescence from the cloud is collected by the microscope objective and imaged onto the Andor EMCCD. Due to the optical density of the in-trap cloud, the probe laser was detuned from the atomic resonance. Optimal fluorescence images were achieved at 3-4 linewidths detuned to the red of the atomic cycling transition.  While this decreases the scattering rate and subsequently the overall signal, it allows the probe light to access a larger fraction of the atoms in the cloud and minimizes displacement of the atoms during the pulse.  Examples of two different optical potentials are shown in Fig.~\ref{fig:SingleBarrier} (c) and Fig.~\ref{fig:MultiBarrier}.  In Fig.~\ref{fig:SingleBarrier} (c) a condensate was split after evaporation by raising a $3~\upmu$K barrier prior to imaging with two probe beams

\noindent  interfering to generate fringes with $5~\upmu$m periodicity. Fig.~\ref{fig:MultiBarrier} (a) and (b) show a 1D array of condensates produced by splitting the cloud prior to the last stage of RF evaporation. The condensates are fully separated by the holographically generated array of barriers, which are shown in Fig.~\ref{fig:MultiBarrier} (c). 

The resolution of the imaging system is characterized by illuminating the atoms with a pair of interfering s-polarized probe beams.  To do so, a second beam offset by an angle θ in the plane of the atom chip is made to interfere with the probe beam shown in Fig.~\ref{fig:Microscope}.  Fig.~\ref{fig:SingleBarrier} (a) shows the results of the interfering probe beams at $\uptheta = 9^{\circ}$, which produces $5~\upmu$m fringes.  This image is a direct summation of five images taken in-trap.  In Fig.~\ref{fig:SingleBarrier} (d) the corresponding Fourier transform of ~\ref{fig:SingleBarrier} (a) is shown, where two peaks resulting from the periodic fringe pattern appear above and below the DC spatial frequency component.  Fig.~\ref{fig:SingleBarrier} (b) and (e) show similar results of the same experiment with $\uptheta = 18^{\circ}$ and the resulting $2.5~\upmu$m fringes.  The small depth of field of the imaging system (about $0.7~\upmu$m) relative to the spatial extent of the cloud (about $9~\upmu$m) limits the effective resolution of the system to about $2.5~\upmu$m.

In conclusion we have demonstrated high-resolution imaging of a Bose-Einstein Condensate through an atom chip with a microscope built from commercially available components.  Additionally, we have shown the ability to simultaneously project optical potentials and image the cloud in-trap. 

We are grateful to L. Czaia for her capable assistance with vacuum cell fabrication. This work was supported in part by the Defense Advanced Research Projects Agency and the Army Research Office (Grant No. W911NF-04-1-0043), the Air Force Office of Scientific Research (Grant No.FA9550-10-1-0135), and the National Science Foundation through a Physics Frontier Center (Grant No. PHY0551010). The work of Evan A. Salim was partially supported by Boulder Labs.

\bibliography{20120820_ImagingAPL}

\end{document}